\documentclass[pre,8.5pt]{article}

\usepackage[super,sort&compress,comma]{natbib} 

\usepackage{sectsty}
\usepackage{balance} 
\usepackage{a4wide}

\usepackage{amsmath}
\usepackage{amsfonts}
\usepackage{amssymb}
\usepackage{verbatim}
\usepackage{graphicx}

\newcommand{\ur}{{\boldsymbol{\hat{u}_r}}}

\newcommand{\utheta}{\boldsymbol{\hat{u}_\theta}}
\newcommand{\uphi}{\boldsymbol{\hat{u}_\varphi}}
\newcommand{\mfric}{{\zeta}_R}

\newcommand{\omeg}{\boldsymbol{{\omega}^\bot}}
\newcommand{\fric}{{\zeta_R}^\bot}
\newcommand{\MSAD}{\langle \Delta \psi^2(t) \rangle}
\newcommand{\Dr}{D_{R}}
\newcommand{\Lm}{L}

\newcommand{\MSD}{\langle \Delta r_{\varphi}^2(t) \rangle}

\begin{document}
\noindent\LARGE{\textbf{3D rotational diffusion microrheology using 2D video microscopy}}
\vspace{0.6cm}

\noindent\large{\textbf{R. Colin, M. Yan, L. Chevry, J.-F. Berret, B. Abou}
\vspace{0.5cm}

\noindent \normalsize{Laboratoire Mati\`ere et Syst\`emes Complexes, UMR CNRS 7057, Universit\'e Paris Diderot\\ 10, rue A. Domon et L. Duquet, 75205 Paris Cedex 13, France}\vspace{0.5cm}

\noindent \normalsize{We propose a simple way to measure the three-dimensional
  rotational diffusion of micrometric wires, using two-dimensional
  video microscopy. The out-of-plane Brownian motion of the wires in a
  viscous fluid is deduced from their projection on the focal plane of
  an optical microscope objective. An angular variable reflecting the
  out-of-plane motion, and satisfying a Langevin equation,
  is computed from the apparent wire length and its projected angular
  displacement. The rotational diffusion coefficient of wires between $1-100\,\mu$m is extracted, as well as the
  diameter distribution. {Translational and rotational diffusion were found to be in good agreement}. This is a promising way to
  characterize soft visco-elastic materials, and probe the dimension
  of anisotropic objects.  }

\section{Introduction}

Recording the three-dimensional (3D) motion of anisotropically shaped
probes is a challenging issue. Although the theory of rotational
Brownian motion has been established for a long time now~\cite{Doi-Edwards},
the direct visualization and quantification of the Brownian motion of
a micrometric anisotropic probe with a microscope is recent. It is due
to the difficulty of quantifying the three-dimensional motion of the
probe with two-dimensional (2D) optical techniques. It was first
solved by studying the in-plane rotational motion of anisotropic
probes~\cite{PRE-Wilhelm-2003,Science-Han-2006}. Recently, highly
specialized optical techniques have opened new
opportunities. Rotational diffusion was studied using light streak
tracking of thin microdisks~\cite{PRL-Cheng-2003,PRE-Wilhelm-2003},
depolarized dynamic light scattering and epifluorescence microscopy of
optically anisotropic spherical colloidal
probes~\cite{PRL-Andablo-2005,Langmuir-Anthony-2006}, scanning
confocal microscopy of colloidal rods with three-dimensional
resolution~\cite{JCIS-Mukhija-2007}, or reconstruction of the wire
position from its hologram observed on the focal plane of a
microscope~\cite{OE-Cheong-2010}. The rotation along the long axis was
also investigated by analyzing the fluorescence images of rodlike
tetramers~\cite{Langmuir-Hong-2006}. Following the 3D rotational
diffusion of an optical probe thus remains costly in equipment, as
well as in computational power.

From a practical point of view, micrometric probes can be used to
determine the relation between stress and deformation in materials
reducing significantly the sample volume, which may be crucial in
biological samples~\cite{JRSI-Abou-2010}. The technique, called
microrheology, is a powerful tool to probe the rheological properties
of complex fluids and biological materials at the micrometric
scale. It can be achieved, either by recording the thermal
fluctuations of probes immersed in the material, or by active
manipulation of the probes~\cite{Review-TWaigh-2005}. While
microrheology based on translational diffusion has been extensively
investigated, the rotational diffusion of anisotropic objects remains
poorly explored. However, it may be of great interest to
investigate the length-scale dependent rheological properties of
heterogeneous structured materials, such as complex fluids or
biological tissues. In the case of anisotropic probes such as wires,
the large aspect ratio of the probe allows for a detectable Brownian
motion, over larger length scales, typically between $1-100\,\mu$m,
than for spherical probes.

In this letter, we propose a simple way to measure the 3D rotational
diffusion of micrometric wires, using 2D video microscopy. The 3D
rotational Brownian motion of the wires immersed in a viscous fluid is
extracted from their 2D projection on the focal plane of a microscope
objective. An angular variable reflecting the out-of-plane motion of
the wires and satisfying a Langevin equation, was computed from the
apparent wire length and its projected angular displacement. The
rotational diffusion coefficient was found to vary over more than $5$
decades, for wires of length between $1-100\,\mu$m and anisotropy
ratios in the range $2 - 2000$. {The resolution of the technique was quantified, by analyzing the wires trajectories.} From these measurements, we were able
to extract the distribution of the wires diameter, in good agreement
with electron microscopy measurements. {Rotational and translational diffusion measurements were compared, giving good agreement \cite{Savin2005, OE-Cheong-2010}.} This provides a new step
towards the reliable use of rotational diffusion to characterize
complex materials with an optical microscope, and probe the dimension
of anisotropic objects.
 
\section{Langevin equations for rotational diffusion}

\begin{figure}
\begin{center}
\includegraphics[width=6.1 cm]{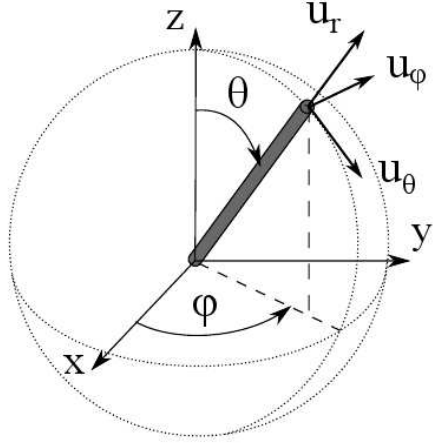}
\caption{\label{convGeom}Spherical coordinate system $(r, \theta, \varphi)$ for a wire
  diffusing in a three-dimensional space. The unit vector $\ur$ is the wire orientational vector.}
\end{center}
\end{figure}

Let us consider a rigid wire diffusing
in a stationary viscous fluid. Its rotational Brownian motion can be modeled
by a Langevin equation, describing the fluctuations of the
wire orientational unit vector $\ur$ (Fig.~\ref{convGeom}). In absence of an
external torque, and neglecting inertia, the rotational equation of motion writes~:
\begin{equation}
0 = -\mfric \boldsymbol{\omega} + \boldsymbol{T_r}\,.
\label{rotationEquation}
\end{equation}
in which $\mfric \boldsymbol{\omega}$ and ${\boldsymbol{T_r}}$ are two
effects from the fluid, respectively the viscous drag and the random
Langevin torque. Since the wire is axisymmetric, the matrix of
friction coefficients $\mfric$ is diagonal in the frame
$(\ur,\utheta,\uphi)$. The eigenvalue along $\ur$ describes the
friction opposing the self-rotation of the wire around its main
axis. The eigenvalues along $\utheta$ and $\uphi$ are equal, and
describe the friction opposing the rotation of the wire main axis.

The projection of the rotational Langevin equation
(\ref{rotationEquation}) on the plane perpendicular to the wire
$(\utheta,\uphi)$, leads to~:
\begin{equation}
-\fric\omeg + \boldsymbol{T_r}^\bot = 0\,,
\label{projectedRotationEquation}
\end{equation}
where $\omeg$ and $\boldsymbol{T_r}^\bot$ are the components, respectively
of the rotation vector and the Langevin random torque, in the plane
$(\utheta,\uphi)$, and $\fric$ is the friction
coefficient perpendicular to the wire axis. The projection of equation (\ref{rotationEquation}) along
the wire axis will not be considered here~\cite{Langmuir-Hong-2006}.

The random torque $\boldsymbol{T_r}^\bot$ can be written as $\boldsymbol{T_r}^\bot
= - T_1\left(t\right)\ \utheta + T_2\left(t\right)\ \uphi$, where
$T_1(t)$ and $T_2(t)$ are two Gaussian white-noise thermal driving torque, satisfying~:
\begin{eqnarray*}
\langle T_{i}\left(t\right)\rangle_{i=1,2}&=& 0\\
\quad\langle T_i(t)
T_j(t')\rangle &= &2\fric k_B \mathrm{T} \delta_{i,j} \delta(t-t')
\label{noise}
\end{eqnarray*}
with $\mathrm{T}$ the bath temperature, and $\langle \, \rangle$ refers
to a time-averaged quantity.

For a finite cylinder -- length $L$ and diameter $d$ -- the
perpendicular friction coefficient can be written in the form~:
\begin{equation}
\fric = \frac{\pi \eta L^3}{3 \,g\left( L/d \right)}
\label{frictionDef}
\end{equation}
where $g\left( L/d \right)$ is a dimensionless function, which takes
into account the finite-size effects of the wire. This function was
analytically calculated for ellipsoids~\cite{perrin-1934}, and
numerically approximated in the case of cylinders~\cite{Doi-Edwards,
  JCP-Tirado-1984, JCP-Broersma-1960-2,
  JCP-Broersma-1981,Fresnais-AdMat-2008}.
 
The rotation vector $\omeg$ can be expressed as $\omeg =
\dot\theta\,\uphi - \sin\theta\,\dot\varphi\,\utheta$, and the
projection of Eq.~\ref{projectedRotationEquation} in the plane
$(\utheta,\uphi)$ thus gives~:
\begin{equation*}
\label{3DLangevin}
\fric \sin\theta\ \dot\varphi = T_1(t), \\
\qquad\fric \dot\theta =
T_2(t)
\end{equation*}
 
The angular variables $\psi(t)$, defined such as $\dot\psi =
\sin\theta\ \dot\varphi$, and $\theta(t)$ both obey a one-dimensional
Langevin equation, which respectively leads to~:
\begin{align}
\langle \Delta\psi^2(t) \rangle = 2 \frac{k_B\mathrm{T}}{\fric} t&= 2 D_R t\label{eqn-psi}\\
\quad\langle \Delta\theta^2(t) \rangle &= 2 D_R t\label{MSADtheory}
\end{align}
where $D_R$ is the rotational diffusion coefficient. Computing Eqs. (\ref{eqn-psi}) and (\ref{frictionDef}), the diffusion coefficient simply writes~:
\begin{equation}
\Dr = \frac{3k_B\mathrm{T}}{\pi\eta \Lm^3} g(\Lm / d)
\label{DiffCoefTheory}
\end{equation}

In the general case of an out-of-plane rotational diffusion,
Eqs. (\ref{eqn-psi}) and (\ref{MSADtheory}) show that determining the
variables $\psi(t)$ or $\theta(t)$ will lead to the rotational
diffusion coefficient $D_R$. From the 2D video recordings, both
$\varphi(t)$ and $\sin\theta(t)$ were extracted. The variable
$\Delta\psi(t) = \psi (t) - \psi (0) = \int_0^t d t'
\sin\theta(t')\ \dot\varphi(t')$ was then computed, leading to the
determination of the mean-squared angular displacement $\langle
\Delta\psi^2(t) \rangle$, and therefore to the rotational diffusion
coefficient $D_R$.

\section{Material and Methods}
\begin{figure}
\begin{center}
\includegraphics[width = 7.3 cm]{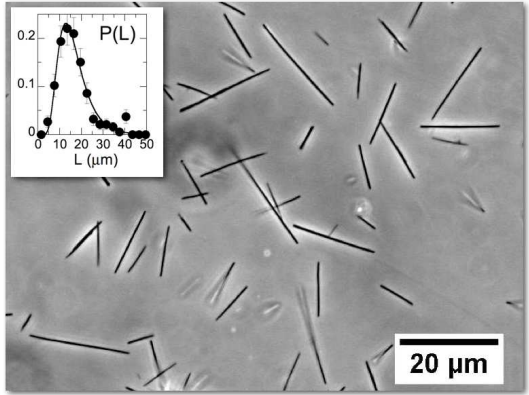}
\caption{Optical microscopy images (X $ 40$) of a wire suspension. The distribution of the lengthes after synthesis can be
  fitted by a log-normal distribution, as shown in inset ($83$
  wires). Due to a rather broad polydispersity, of the order of $0.5$
  (defined as the standard deviation of $\ln{L}$), wires of length between 
  $1$ to $100 \,\mu$m were obtained. The distribution of the diameters was determined from electron microscopy, with a median
  diameter $0.4 \,\mu$m, leading to anisotropy ratios between
  $2-2000$. }
\label{distribL}
\end{center}
\end{figure}

The wire formation results from the electrostatic co-assembly between
oppositely charged iron oxide nanoparticles and
polymers~\cite{Fresnais-AdMat-2008}. The
wires are purified and suspended in DI
water. Figure \ref{distribL} shows an optical microscopy image of the
wire suspension after synthesis, and the corresponding length
distribution in inset. In the present study, two batches of wires of
length $15 \,\mu$m (Fig.~\ref{distribL}) and $30 \, \mu$m are
investigated~\cite{2010-SoftMatt-Minhao}. Due to a rather broad
polydispersity in length, wires of length between $1$ and $100
\,\mu$m were obtained. The distribution of the wires diameter was
determined from electron microscopy, with a median diameter $0.4
\,\mu$m, leading to anisotropy ratios between $2-2000$.

The aqueous wire suspension was then mixed with pure glycerol
giving aqueous solutions of glycerol, also referred as wire
suspensions. Aqueous solutions of glycerol with two different volume
fractions, $50\%$ and $60\%$, were prepared. The wire suspension was
then introduced in an observation chamber ($3\,\rm{mm} \times 3\,\rm{mm}
\times 250 \,\rm{\mu m}$) between a microscope slide and a coverslip,
sealed with araldite glue to avoid evaporation and contamination of
the sample.

An inverted Leica DM IRB microscope with a $\times100$ oil immersion
objective (NA=1.3, free working distance : $130\, \mu$m), coupled to a camera (EoSens Mikrotron) were used to record
the 2D projection of the wires thermal fluctuations on the focal plane
objective. The wires concentration was chosen diluted enough to
prevent collisions and hydrodynamic coupling. They were always tracked far
enough from the walls of the observation chamber. The microscope objective
temperature was controlled within $0.1\,{^\circ}$~C, using a Bioptechs
heating ring coupled to a home-made cooling device. The sample
temperature was controlled through the oil immersion in
contact. Sedimentation of the wires was negligible on the recording
time scales.

The camera was typically recording $10$ images per second during
$200$~s ($2000$ images). The 3D Brownian motion of the wires was
extracted from their 2D projection on the $(x,y)$ plane
(Fig.~\ref{convGeom}). The angle $\varphi(t)$ and the projected length
$L(t)$ were measured from the images, using a home-made tracking
algorithm, which is implemented as an ImageJ
plugin\cite{ImageJ}. Since the objective depth of focus ($1-2\,\mu$m)
is shorter than the length of the wires, their out-of-plane image is
distorted. The distorted image of a wire was seen as a cluster of
connected pixels. The algorithm output is the length $L(t)$ of the
cluster (projected length of the wire) and the angle $\varphi(t)$ of
the cluster, as defined in figure \ref{convGeom}. The algorithm mainly
consists in three steps. First, a user-defined threshold is applied to
the image. Then, the wire projection, seen as a set of connected
pixels (cluster) is tracked at time $t$, in the vicinity of the
position at time $t-1$. Finally, the cluster orientation and its
length are computed, respectively giving the angle $\varphi(t)$, and
the projected length $L(t)$. The position of the cluster center was
also computed. Wires with a high out-of-plane angle, corresponding to
angle $\theta$ smaller than $50$ degrees, could not be
considered. Since the tracked wires are chosen to lie in the focal
plane at the beginning of the recording, the length of the wire $L$
was taken as the maximum measured length $L(t)$ within the recording
time, leading to $\sin\theta= L(t)/\max{(L(t))}$. The quantity
$\Delta\psi(t)$ was computed by using the discrete equation $
\Delta\psi(t, t_0) = \sum_{t'=t_0}^{t_0+t-\delta t} \frac{L(t')}{\Lm}
\delta\varphi(t')$, where $\delta\varphi(t') = \varphi(t'+\delta t) -
\varphi(t')$, and $\delta t$ is the time lapse between two images. A
time average then enables us to compute $\langle \Delta\psi^2(t)
\rangle = \langle \left(\Delta\psi^2(t, t_0)\right) \rangle_{t_0}$.

{The  uncertainty on $\MSAD$ was evaluated including the uncertainties
  on $\varphi$, the projected length, and the total length. It also takes into account statistical
  accuracy. The uncertainty on the apparent length of an out-of-plane
  wire was determined by varying the $z$-position of the focal plane
  while recording the fixed wire. The corresponding computed lengths for different
  $z$-positions of the focal plane give an uncertainty of the apparent
  length as the wire moves along the axial direction. It was estimated
  to be $8 \%$ for wires longer than $4 \, \mu$m and $15\%$ below. }

\section{Rotational diffusion coefficient and diameters distribution}
\begin{figure}
\begin{center}
\includegraphics[width = 7.9 cm]{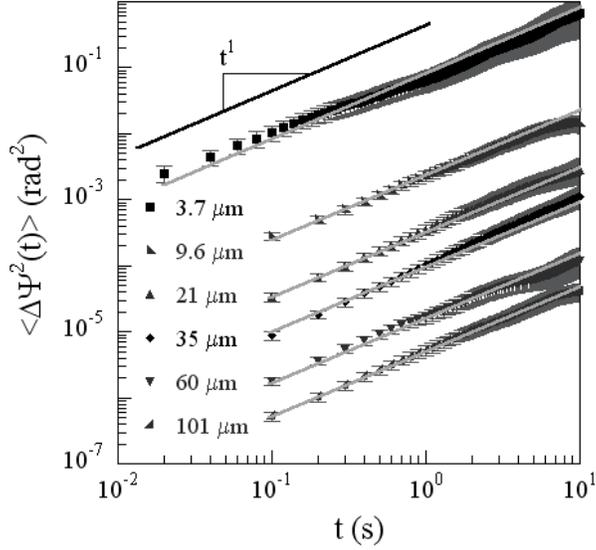}
\caption{Mean-squared angular displacement (MSAD) of the wires as a
  function of the lag time, for wires of length between $3$ and
  $100\,\mu$m. The experiments were performed in an aqueous solution
  of glycerol ($50\%$ in volume). The MSAD increases linearly with
  time, as expected in a purely viscous fluid. The slope of each
  curve, corresponding to the diffusion coefficient, decreases with
  the length of the wire. Similar data were obtained in the $60\%$ in
  volume aqueous solutions of glycerol (not shown). The error bars
  include the uncertainties on $L$, $\varphi$ and statistical
  accuracy. }
\label{MSADmes}
\end{center}
\end{figure}

\begin{figure}
\begin{center}
\includegraphics[bb=14 14 479 450,clip,width = 8.6 cm]{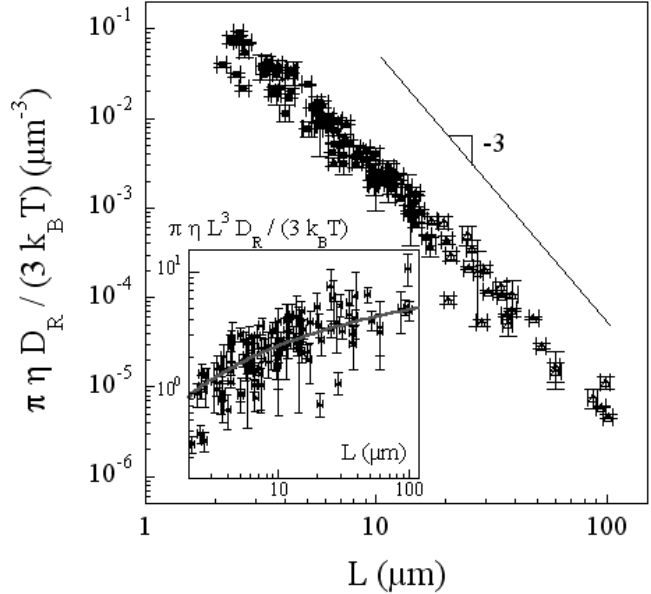}
\caption{Rotational diffusion coefficient of the wires as a function
  of the wires length, normalised by viscosity and temperature. At first order, the diffusion coefficient scales as
  $\Lm^{-3}$, as expected from Eq.~(\ref{DiffCoefTheory}). The
  dimensionless parameter $g_{rot}(\Lm /d)=\frac{\pi \eta
    \Lm^3}{3k_B\mathrm{T}}\Dr$ is shown in inset as a function of the
  wires length. The gray line corresponds to the analytical
  expression established by Broersma, assuming a median diameter of
  $400$ nm. The large distribution of the data around the gray line
  reflects the distribution of the wires diameter.}
\label{diffCoefMes}
\end{center}
\end{figure}

The mean-squared angular displacement (MSAD) $\MSAD$ is shown in
Fig.~\ref{MSADmes} as a function of the lag time $t$, for wires of
length between $3\, \rm{\mu }$m and $100\, \rm{\mu }$m. The MSAD
$\MSAD$ was found to increase linearly with time, as expected in a
viscous fluid. From these curves, a rotational diffusion coefficient
$\Dr$ defined such as $\MSAD= 2 \Dr t$, could then be
extracted. Figure~\ref{diffCoefMes} shows the rescaled quantity
$\frac{\pi \eta}{3k_B\mathrm{T}}\Dr$ as a function of the measured
length of the wire $L$. At leading order, the diffusion coefficient
decreases as $\Lm^{-3}$, over $5$ decades. A correction $g_{rot}(\Lm / d)$
for the finite-size effects of the wire is expected, as described in
Eq.~(\ref{DiffCoefTheory}). The rescaled diffusion coefficient
$\frac{\pi \eta}{3k_B\mathrm{T}}\Dr$ was then multiplied by $\Lm^{3}$,
leading to the experimental determination of the dimensionless
function $g_{rot}(\Lm / d) $ (Fig.~\ref{diffCoefMes}-inset). 

{The resolution of the technique was quantified from the wires
  trajectories using the relation :
\begin{equation}
\MSAD= 2 D_R (t- \sigma/3) +
  2\epsilon_{rot}^2
\label{error-rot-eqn}
\end{equation}
including the measurement error $\epsilon_{rot}$,
  and corrected for the camera exposure time $\sigma$
  \cite{Savin2005,OE-Cheong-2010}. The measurement error
  $\epsilon_{rot}$ was found to decrease with the length of the wire, as shown in figure \ref{error-rot}. It typically
  corresponds to an error in orientation of $4 ^\circ$ for a $10
  \,\mu$m long wire. The largest error was obtained for short wires,
  of $2$ to $3$ micrometers long, where an error of orientation of roughly
  $ 8^\circ$ was found. For the longest wires above $20
  \,\mu$m, the error was found
  to be less than $1  ^\circ$. 
\begin{figure}
\begin{center}
\includegraphics[width = 8.0 cm]{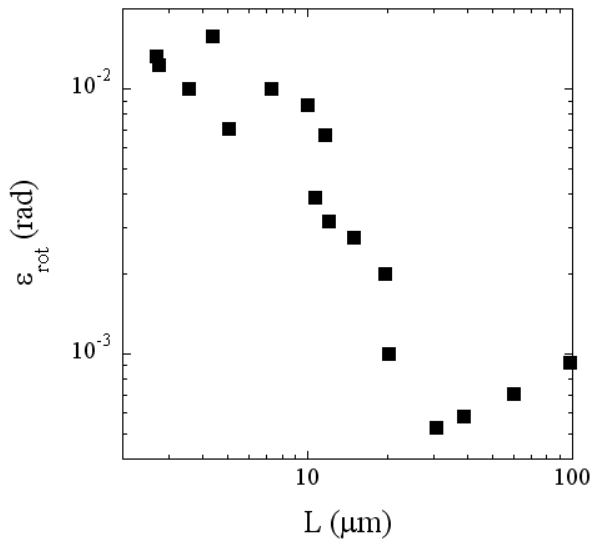}
\caption{Measurement error $\epsilon_{rot}$ as a function of the length of the wires. The error in orientation is around $4 ^\circ$ for a $10
  \,\mu$m long wire. The largest error was obtained for short wires,
  of $2$ to $3$ micrometers long, where an error of orientation of roughly
  $ 8^\circ$ was found. For the longest wires above $20
  \,\mu$m, the error was found
  to be less than $1 ^\circ$. }
\label{error-rot}
\end{center}
\end{figure}
}

An analytical expression for the finite-size effects function $g_{rot}(\Lm /
d)$ was established by Broersma in 1960~\cite{JCP-Broersma-1960,
  JCP-Broersma-1981}. It is expected to be valid for $p = \Lm / d >
4.6$, and writes~:
\begin{eqnarray}
g_{rot}(p) = \ln\left(p\right) - 0.446 - \frac{0.2}{\ln\left(2p\right)} \nonumber\\
-
\frac{16}{\ln\left(2p\right)^2}
 + \frac{63}{\ln\left(2p\right)^3} -
\frac{64}{\ln\left(2p\right)^4}
\label{broersma}
\end{eqnarray}
Fig.~\ref{diffCoefMes}-inset shows the Broesma relation~(\ref{broersma}),
using a median diameter $d= 400 $~nm. The large distribution of
  the data around this adjustment reflects the distribution of
the diameter of the wires. 

By numerically inverting the
relation~(\ref{broersma}), we could extract the distribution of the wires diameter. The corresponding
values of $p=L/d$ range between $2$ and $2000$. A
small fraction of the wires (less than $5\%$) exhibit a value of $p$
which is smaller than the validity range established by Broersma
($p>4.6$). More recently, another analytical expression, valid for $2
< p< 20$, was proposed by Tirado \emph{et al.}~\cite{JCP-Tirado-1984}. This expression was used for the few wires falling out of the validity range of
Broersma's theory, and for which Tirado's expression is valid. 

Figure~\ref{diam-distr} shows the distribution of the diameters obtained from
the diffusion measurements, also compared to the one obtained
from Scanning Electron Microscopy (SEM) measurements. The
distributions have been fitted by log-normal distributions, leading
to equal median diameters $ 0.4\,\rm{\mu }$m. The polydispersities
(standard deviations of $\ln d$) are respectively ${\sigma_d}_{\rm
  micro} = 0.53 $ and ${\sigma_d}_{\rm SEM} = 0.38$. The distributions
are in good agreement, the largest discrepencies come for the smallest
diameters, where the precision of our method is the lowest, because of
the high non-linearity of the function $g_{rot}(L/d)$ as a function of $d$.

\begin{figure}
\begin{center}
\includegraphics[width = 8.0 cm]{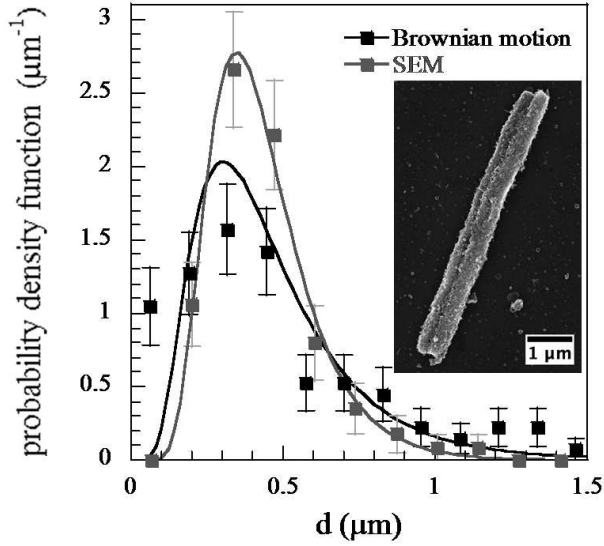}
\caption{Distribution of the wires diameter, extracted from the wires Brownian motion measurements. The
  distributions are fitted by log-normal distributions, leading to equal median
  diameters $0.4\, \rm{\mu }$m. The polydispersities are respectively ${\sigma_d}_{\rm micro} = 0.53 $ and ${\sigma_d}_{\rm SEM} =  0.38$. 
The distributions, obtained with differents methods are in good agreement. In inset, scanning electron microscopy of a wire.  }
\label{diam-distr}
\end{center}
\end{figure}

{We now compare rotational and translational fluctuations. The translational fluctuations
  of the wires were measured in the wire's frame of reference. In our
  experiments, only the component of the center-of-mass translation
  along $\uphi$ could be measured. Since the diffusion in the
  $z$-direction could not be evaluated, the translational diffusion
  parallel to the wire, along $\ur$, and the one along $\utheta$ were
  not accessible. Figure \ref{trans} shows the center-of-mass
  mean-squared displacement $\MSD=\langle ((\boldsymbol{r}(t+t')-\boldsymbol{r}(t')) \cdot \uphi(t'))^2\rangle_{t'}$ for  wires, between $3$ and $ 100
  \, \mu$m. It was found to increase linearly with the lag time, as
  expected in a viscous fluid. The data of translational diffusion
  were fitted according to the relation \cite{Savin2005}:
\begin{eqnarray}
\MSD = 2 D_{transl}^{\perp}(t-\sigma/3) + 2\epsilon_{transl}^2
\label{error-trans}
\end{eqnarray} 
including
  the measurement error $\epsilon_{transl}$ and corrected for the
  camera's exposure time $\sigma$. The translational diffusion
  coefficient writes :
 \begin{equation}
D_{transl}^{\perp}= \frac{k_BT}{ 4 \pi \eta L}
  g_{transl}^{\perp}(p)
\label{D-trans}
\end{equation}
with the finite-size effect function $
  g_{transl}^{\perp}(p)= ln(p) + 0.839 + 0.185/p +
  0.233/p^2$\cite{JCP-Tirado-1984}. The measurement error $\epsilon_{transl}$ is shown in
  Figure \ref{trans} (inset) for wires of length between $3$ and $100\,\mu$m. It was
  found to be less than $\epsilon_{transl}= 0.3$ pixel in most
  cases, which means that the center of the wire was tracked with a
  precision better than $40$ nm in the $(x,y)$ plane.

Fitting Eqs (\ref{error-rot-eqn}) and (\ref{error-trans}) for a given
wire gives the values $D_{R}$ and $D_{transl}^{\perp}$ of the
diffusion coefficients, respectively obtained with rotational and
translational measurements. Given the experimental conditions, this
respectively yields values of $\ln(p)$, $\ln(p_{rot})$ and
$\ln(p_{transl})$, according to Eqs. (\ref{DiffCoefTheory}) and
(\ref{D-trans}). Table 1 presents a comparison between rotational and
translational measurements for wires of length between $4$ and $95
\,\mu$m. The values of $\ln(p)$ obtained with rotational and
translational measurements were found to be in good agreement.

\begin{table*}
\begin{center}
\begin{tabular}{|c|c|c|c|c|}
 \hline 
      & & & & \\
 $L (\mu\rm{m})$ & $D_{R} \,(\rm{rad}^2/$s) & $10^3 D_{transl}^{\perp} \,(\mu\rm{m}^2/$s) & $\ln(p_{rot})$ & $\ln(p_{transl})$\\ 
 & & & & \\
 \hline 
 $4.4 $ & $0.016 \pm 0.004$  & $39 \pm 4 $ & $3.3\pm 1.1 $ & $3.4\pm 0.6$ \\ 
 \hline
 $9.8$ & $(1.3\pm0.2)10^{-3} $ & $17.9\pm0.7 $ & $3.2\pm0.7 $ & $3.6 \pm0.3$ \\
 \hline 
 $35$ & $(4.9 \pm 0.2)10^{-5} $ & $6.7\pm 0.2$ & $5.2\pm 0.7$ & $5.2 \pm 0.2$ \\ 
 \hline 
 $95$ & $(2.8 \pm0.1) 10^{-6} $ & $3.1 \pm 0.2 $ & $5.8 \pm 0.7$ & $6.8 \pm 0.5$\\ 
\hline
\end{tabular}
\label{table1}
\caption{Comparison between rotational and translational
  measurements for wires of various lengths. The values of $\ln(p)$ obtained with rotational and translational measurements, respectively $\ln(p_{rot})$ and $\ln(p_{transl})$, are in good agreement. }
\end{center}
\end{table*}

\begin{figure}
\begin{center}
\includegraphics[width = 8.0 cm]{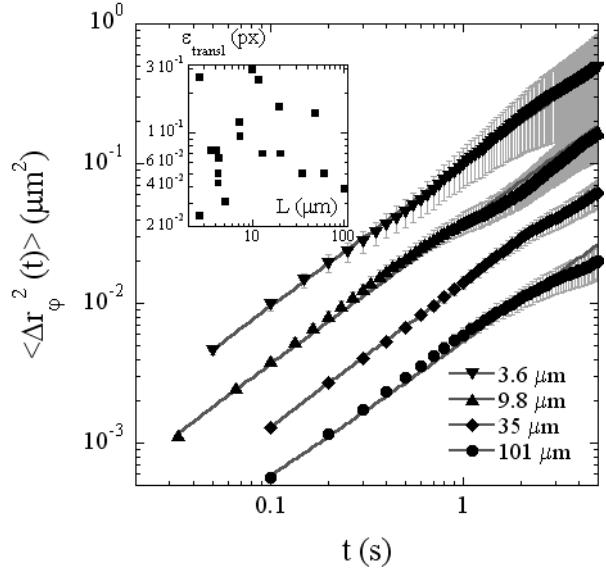}
\caption{Mean-squared displacement of the center-of-mass of the wires of length between $3$ and $ 100\,\mu$m. The MSD increases linearly with time, as expected in a purely viscous fluid. The lines correspond to the fits according to Eq. (\ref{error-trans}). Inset: measurement error $\epsilon_{transl}$ for wires of various lengths.
}
\label{trans}
\end{center}
\end{figure}
}

\section{Conclusion}

In this letter, we propose a simple way to follow the 3D rotational
Brownian motion of micrometric wires in a viscous fluid, from the 2D
projection of the wires on the focal plane of a microscope. The
rotational diffusion coefficient of the wires between $1-100 ~\mu$m
was computed as a function of the wires length, ranging over $5$
decades. {The resolution of the technique was quantified, by
  analyzing the wires trajectories}. Our diffusion measurements allow us to extract the
distribution of the wires diameter, in good agreement with SEM
measurements. {Rotational and translational  diffusion measurements were compared and found to be in good agreement \cite{Savin2005,
    OE-Cheong-2010}.} This technique provides a simple way to measure
the out-of-plane rotational diffusion of a wire in a viscous fluid,
opening new opportunities in microrheology to characterize more
complex fluids, and probe the dimension of anisotropic objects.

\section{Acknowledgements}
We thank O. Sandre and J. Fresnais from the Laboratoire Physico-chimie
des Electrolytes, Collo\"{\i}des et Sciences Analytiques (UMR CNRS
7612) for providing us with the magnetic nanoparticles. This research
was supported by the ANR (ANR-09-NANO-P200-36) and the European
Community through the project NANO3T (number 214137
(FP7-NMP-2007-SMALL-1).

\bibliographystyle{unsrt} 
\bibliography{bibliorod2011-12-13}

\end{document}